\documentstyle[preprint,prd,aps]{revtex}

\def\gsim{\begin{array}{c} > \\ \sim \end{array}}
\def\lsim{\begin{array}{c} < \\ \sim \end{array}}
\begin{document}
\draft
\title{\large \bf Measuring the Foaminess of Space-Time with Gravity-Wave
Interferometers}
\author{\bf Y. Jack Ng\thanks{e-mail: yjng@physics.unc.edu} and H. van
Dam}
\address{Institute of Field Physics, Department of Physics and Astronomy,
University of North Carolina, Chapel Hill, NC 27599-3255}
\maketitle
\bigskip
\begin{verse}
\hspace{0.7in}{\em 'T was noted in heaven, 't was felt in hell,} \\
\hspace{0.7in}{\em And echo caught faintly the noise as it fell...} \\
\hspace{0.9in}(Slightly modified from {\sl "Enigma. The letter H"} by C.M. 
Fanshawe)
\end{verse}

\begin{abstract}
By analyzing a gedanken experiment designed to measure the distance $l$
between two
spatially separated points, we find that this distance cannot be measured
with uncertainty less than $(ll_P^2)^{1/3}$, considerably larger
than the Planck scale $l_P$ (or the string scale in string theories), the
conventional-wisdom uncertainty in distance measurements.
This limitation to space-time measurements is interpreted as resulting from 
quantum fluctuations of space-time itself.  Thus, at very short
distance scales, space-time is "foamy."  This intrinsic foaminess of
space-time provides
another source of noise in the interferometers.  The LIGO/VIRGO and
LISA generations of gravity-wave interferometers, through future
refinements, are expected to reach
displacement noise levels low enough to test our proposed degree of foaminess 
in the structure of space-time.  We also point out a simple connection to
the holographic principle which asserts that the number of degrees of
freedom of a region of space is bounded by the area of the region in Planck
units.
\end{abstract}

\newpage
\section{Introduction}

Quantum mechanics and general relativity, the two pillars of modern
physics, are very useful in describing the phenomena 
in their respective domains of physics.  Unfortunately, their
synthesis has been considerably less successful.  It is better known for
producing a plethora of puzzles from the embarrassing cosmological
constant problem\cite{ccp} to the enigma of possible information loss
associated with black hole evaporation\cite{bhe}.  String theory is a
reaction to this crisis.  Nowadays, it is the main contender to be the
microscopic theory of quantum gravity.  But even without the correct
theory of quantum gravity (be it string
theory or something else), we know enough about quantum mechanics and
gravity to study its low-energy limit.  In particular, we would like to
know what that limit of quantum gravity can tell us about the structure of
space-time.  In this article, we will combine the general
principles of quantum mechanics with those of general relativity to
address the problem of quantum measurements of space-time distances.

But first, let us recall what quantum mechanics and general relativity
have to say about the nature of space-time distance measurements.  In
quantum mechanics, we specify a space-time point simply by its
coordinates; hardly do we feel the need to give a prescription to spell  
out how the coordinates are to be measured.  This lax attitude will not do
with general relativity.  According to general relativity, coordinates do
not have any meaning independent of observations; in fact, a coordinate
system is defined only by explicitly carrying out space-time distance
measurements.  In the following (the discussion is based on our earlier 
work\cite{nvd1,nvd2})
we will abide by this rule of general relativity, and will follow
Wigner\cite{wigner} in using clocks and light signals to measure
space-time distances.

In Section II, we will analyze a gedanken experiment designed to measure
the distance between two spatially separated points, and will show that
quantum mechanics and general relativity together imply that there is a
limit on the accuracy with which we can measure that distance.  That
uncertainty in space-time measurements is interpreted to induce an uncertainty in the
space-time metrics; in other words, space-time undergoes quantum
fluctuations.  Some consequences of space-time fluctuations are listed
in Section III.  Section IV is devoted to show how gravity-wave 
interferometers can be
used to test this phenomenon of space-time fluctuations.  We offer our
conclusions in Section V.

\section{From Space-time Measurements to Space-time Foams}

Suppose we want to measure the distance between two separated points A and
B.  To do this, we put a clock (which also serves as a light-emitter and
receiver) at A and a mirror at B.  A light signal is sent from A to B
where it is reflected to return to A.  
If the clock reads zero when the light signal is emitted and reads $t$
when the signal returns to A, then
the distance between A and B is given by $l = ct/2$, where $c$
stands
 for the speed of light.  The next question is: What is the
uncertainty (or error) in the distance measurement?  Since the clock at A
and the mirror at B are the agents in measuring the distance, the
uncertainty of distance $l$ is given by the uncertainties in their
positions.  We will concentrate on the clock, expecting that the mirror
contributes a comparable amount to the uncertainty in the measurement of $l$.
Let us first recall that the clock is not stationary; its spread in speed at 
time zero is given by
the Heisenberg uncertainty principle as
\begin{equation}
\delta v =  \frac{\delta p}{m} \gsim \frac{\hbar}{2m\delta l},
\label{ineq1}
\end{equation}
where $m$ is the mass of the clock.  This implies an uncertainty in the
distance at time $t$,
\begin{equation}
\delta l(t) = t \delta v \gsim \left(\frac{\hbar}{m \delta l(0)}\right)
\left(\frac{l}{c}\right), 
\label{ineq2}
\end{equation}
where we have used $t/2 = l/c$ (and we have dropped an additive term
$\delta l(0)$ from the right hand side since its presence complicates the
algebra but does not change any of the results). Minimizing $(\delta l(0)
+ \delta l(t))/2$ we get \begin{equation}
\delta l^2 \gsim \frac{\hbar l}{mc}
\label{ineq3}
\end{equation}

At first sight, it appears that we can make $\delta l$, the uncertainty in
the position of the clock, arbitrarily small by using a clock with
a large enough (inertial) mass.
But that is wrong as the (gravitational) mass of the clock would disturb
the curvature.  It is here the principle of equivalence in general
relativity comes into play: one cannot have a large inertial mass and a
small gravitional mass since they are equal.  We can now exploit this
equality of the two masses to eliminate the 
dependence on $m$ in the above inequality to make the uncertainty
expression useful.  Let the clock at A be a
light-clock consisting of two parallel mirrors (each of mass $m/2$), a
distance of $d$ apart, between which bounces a beam of light.  On the one
hand, the clock must tick off time fast enough such that $d/c < \delta
l/c$, in order that the distance uncertainty is not greater than $\delta l$.
On the other hand, $d$ is necessarily larger than the Schwarzschild radius
$Gm/c^2$ of the mirrors ($G$ is Newton's constant) so that the time
registered by the clock can be read off at all.  From these two
requirements, it follows that
\begin{equation}
\delta l > d > \frac{Gm}{c^2},
\label{ineq4}
\end{equation}
the product of which and Eq. (\ref{ineq3}) yields \cite{karol} 
\begin{equation}
\delta l \gsim (l l_P^2)^{1/3},
\label{ineq5}
\end{equation}
where $l_P = (\frac{\hbar G}{c^3})^{1/2}$ is the Planck length ($\sim
10^{-33}$ cm).  In a
similar way, we can
deduce the uncertainty in time interval ($t$) measurements,
\begin{equation}
\delta t \gsim (t t_P^2)^{1/3},
\label{ineq6}
\end{equation}
where $t_P = l_P/c$ is the Planck time ($\sim 10^{-42}$ sec).  

The intrinsic uncertainty in space-time measurements just described can be interpreted as inducing
an intrinsic uncertainty in the space-time metric $g_{\mu \nu}$.
Noting that $\delta l^2 = l^2 \delta g$ and using Eq. (\ref{ineq5}) we get
\begin{equation}
\delta g_{\mu \nu} \gsim (l_P/l)^{2/3} \sim (t_P/t)^{2/3}.
\label{delg}
\end{equation}
The fact that there is an uncertainty in the space-time metric means that
space-time is foamy.  The origin of the uncertainty is quantum mechanical.
Therefore we can say that space-time undergoes quantum fluctuations and
this is an intrinsic property of space-time.  The amount of fluctuations
on a length scale $l$ or time scale $t$ is given by Eq. (\ref{delg}).

The uncertainty expressed in Eq. (\ref{ineq3}) is due to quantum effects, 
and it depends on $m$, the mass of the clock.  In the above, we have used
Eq. (\ref{ineq4}) to put a bound on $m$, eventually arriving at Eq.
(\ref{ineq5}).  Perhaps, we should point out that, besides Eq.
(\ref{ineq5}), there are (at least) two other expressions for the 
uncertainty in space-time measurements that have appeared in the
literature, predicting different degrees of foaminess in the structure of 
space-time.  Instead of repeating the derivations used by the other
workers, we find it instructive to "derive" them by adopting an argument
similar to the one we have used above.  We start with Eq. (\ref{ineq3}).
For the bound on $m$, if one uses (instead of Eq. (\ref{ineq4}))
\begin{equation}
l \gsim \frac{Gm}{c^2},
\label{cons}
\end{equation}
then one finds
\begin{equation}
\delta l \gsim l_P,
\label{Wheeler}
\end{equation}
the canonical uncertainty in distance measurements widely quoted in the
literature\cite{MTW}.  Eq. (\ref{cons}) gives a considerably more
conservative bound on $m$; and the inequality is trivially satisfied
because, otherwise, point B would be inside the Schwarzschild radius
of the
clock at A, an obviously nonsensical situation.  So, we do not expect the
resulting inequality (given by Eq. (\ref{Wheeler})) to be very restrictive
(or, for that matter, to be very useful, in our opinion).  

On the other hand, if, instead of Eq. (\ref{ineq4}), one uses 
\begin{equation}
m_P \gsim m,
\label{radi}
\end{equation}
where $m_P \equiv \hbar / c l_P$ denotes the Planck mass ($\sim
10^{-5}$ gm),
then combining it with Eq. (\ref{ineq3}), one gets
\begin{equation}
\delta l \gsim (l l_P)^{1/2},
\label{AC}
\end{equation}
a result for the uncertainty in space-time measurements found in
Ref. \cite{GAC}.  Since $l >> l_P$ (which we have implicitly assumed), the
distance uncertainty given by Eq. (\ref{AC}) is considerably bigger than the
one proposed by us (Eq. (\ref{ineq5})).  But regardless which of the
three pictures of
space-time foam we have in mind, they all predict a very small distance
uncertainty: e.g., even on the
size of the whole observable universe ($\sim 10^{10}$ light-years), Eq.
(\ref{Wheeler}), Eq. (\ref{ineq5}), and Eq. (\ref{AC}) yield a fluctuation
of only about $10^{-35}$ m, $10^{-15}$ m and $10^{-4}$ m respectively.  We
leave it to the readers
to decide for themselves which of the three pictures of space-time foam is
the most reasonable.
  
\section{Other Properties of Space-time Foam}

Let us return to that picture of space-time foam proposed by us, expressed
in Eq. (\ref{ineq5}), Eq. (\ref{ineq6}), and Eq. (\ref{delg}).  The metric
fluctuations give rise to some rather interesting properties
\cite{nvd1,nvd2} besides the uncertainties in space-time measurements.  
Here is a partial list:

(i)  There is a corresponding uncertainty in energy-momentum
measurements for elementary particles, given by
\begin{equation}
\delta p \gsim p \left(\frac{p}{m_P c}\right)^{2/3}, 
\hspace{0.5in} \delta E \gsim E \left(\frac{E}{m_P
c^2}\right)^{2/3}.
\label{p&E}
\end{equation}
We should keep in mind that energy-momentum is conserved only up to this
uncertainty.

(ii)  Space-time fluctuations lead to decoherence phenomena.  The point
is that the metric fluctuation $\delta g$ induces a multiplicative phase 
factor in the wave-function of a particle (of mass $m$)
\begin{equation}
\psi \rightarrow e^{i \delta \phi} \psi,
\label{wf}
\end{equation}
given by
\begin{equation}
\delta \phi = \frac{1}{\hbar} \int m c^2 \delta g^{00} dt.
\label{phase}
\end{equation}
One consequence of this additonal phase is that a point particle with mass 
$m > m_P$ is a
classical particle (i.e., it suffices to treat it classically).  This fuels 
the speculation that the high energy
limit of quantum gravity is actually classical.  But in connection with
this speculation, a cautionary remark is in order: by extrapolating 
the mass scale beyond the Planck mass, one runs the risk of going
beyond the domain of validity of this
work, viz. the low-energy limit of quantum gravity.\cite{twoforms}

(iii)  The energy density $\rho$ associated with the metric fluctuations
(Eq. (\ref{delg})) is actually very small.  
Regarding the metric fluctuation as a
gravitational wave quantized in a spatial box of volume $V$, we find
\begin{equation}
\rho \sim m_P c^2 / V.
\label{rho}
\end{equation}
However, if one uses the "root mean square" approach proposed in the first
paper in Ref. \cite{karol}, one gets an unacceptably large 
energy density of $m_P c^2 / l_P^3$.

(iv)  Due to space-time fluctuations, gravitational fields of individual
particles with mass $m << m_P$ that make up ordinary matter are not
observable.  From this point of view, the gravitational field is a
statistical phenomenon of bulk matter.\cite{dewitt}

(v)  There is a simple connection between spacetime quantum fluctuations as
given by Eq. (\ref{ineq5}) and the holographic principle\cite{thooft}.  The 
holographic principle asserts that the number of degrees of freedom of a 
region of space is bounded by the area of the region in Planck units.  To see the connection, consider a region of space with linear dimension $l$.  
According to the conventional wisdom, the region can be partitioned into cubes
as small as $l_P^3$.  It follows that the number of degrees of freedom of the 
region is bounded by $(l/l_P)^3$, i.e., the volume of the region in Planck units.  But according to our spacetime foam picture (Eq. (\ref{ineq5})), the 
smallest cubes inside that region have a linear dimension of order 
$(l l_{P}^2)^{1/3}$.  Accordingly, the number of degrees of freedom of the region is bounded by $[l/(ll_{P}^2)^{1/3}]^3$, i.e., the area of the region in
Planck units, as stipulated by the holographic principle.  Thus one may say
that the holographic principle has its origin in the quantum fluctuations of 
spacetime.  It has not escaped our attention that
the effective dimensional reduction of the number of degrees of 
freedom may have a dramatic effect on the ultraviolet behaviour of a quantum
field theory.

(vi)  Fluctuations in space-time imply that metrics can be defined only
as averages over local regions and cannot have meaning locally.  This
gives rise to some sort of non-locality.  It has also been
observed\cite{ahlu} that the space-time measurements described above alter
the space-time metric in a fundamental manner and that this unavoidable
change in the metric destroys the commutativity (and hence locality) of
position measurement operators.  The gravitationally induced non-locality,
in turn, suggests a modification of the fundamental commutators.
Furthermore, we would not be surprised 
if this feature of non-locality is in some way related to the
holographic principle\cite{thooft}. 
 
\section{Probing the Structure of Space-time with Gravity-wave
Interferometers}

As noted above, the fluctuations that space-time undergoes are extremely
small.  Indeed, it is generally believed that no currently
available technologies are powerful enough to probe into the 
space-time foam.  But it has been shown\cite{amca} recently by G.
Amelino-Camelia that modern gravity-wave interferometers are
already sensitive enough or will soon be sensitive enough to test two of
the three pictures of space-time foam described in Section II. 

First let us briefly recall the physics of modern gravity-wave 
interferometers.  They consist of a laser light source, a beam splitter,
and two mirrors placed at the ends of two (very long) arms arranged in an
L-shaped
pattern.  The light beam is split by the beam splitter into a transmitted
beam and a reflected beam.  The transmitted beam is directed toward one of
the mirrors; and the reflected
beam is directed toward the other mirror.  The two beams of light are
reflected by the mirrors back to the beam splitter where they are
superposed.  The resulting interference pattern is very sensitive to
changes in the distances between the beam splitter and the mirrors at the
ends of each arm.  Modern
gravity-wave interferometers are sensitive to changes in distance to an
accuracy of the order of $10^{-18} m$ and better.  To reach such a
sensitivity, one has to contend with all sorts of noises such as seismic
noise, suspension thermal noise, and photon shot noise.  Our claim is that
even after one has subtracted away all these known noises, there is still
a noise arising from space-time fluctuations.

At first sight, it appears that the task of measuring space-time
fluctuations is well beyond our reach; after all, even the extraordinary
sensitivity down to an accuracy of order $10^{-18}$ m is no where
near the Planck scale of $10^{-35}$ m.  But 
the displacement sensitivity of an interferometer actually
depends on frequencies $f$ (more on this below).  Besides the $10^{-18}$ 
m length scale mentioned above, the physics of
interferometers involves another length scale $c/f$ provided by $f$.
Interestingly, as shown in Ref. \cite{amca}, within certain range of
frequencies, the experimental
limits are comparable to the theoretical predictions for two of the
space-time foam pictures described above.   

The idea of using gravity-wave interferometers to probe the structure of
space-time is actually fairly simple.  Let us concentrate on the picture
of space-time foam described by Eq. (\ref{delg}) and Eq. (\ref{ineq5}) or
Eq. (\ref{ineq6}).  Due to the foaminess of space-time, in any distance
measurement that involves an amount of time $t$, there is a minute
uncertainty $\delta l \sim (ct l_{QG}^2)^{1/3}$, where, for later use,  
we have introduced $l_{QG}$ which we expect to be of order $l_P$.
(It is understood that the time of observation $t$ is much smaller than the 
time interval over which the space-time region where the observation is done 
experiences significant curvature effects.)
But measuring minute changes in (the) relative distances (of the test
masses or the mirrors) is exactly what an interferometer is designed to
do.  Hence, the intrinsic uncertainty in a distance measurement for a time
$t$ manifests itself as a displacement noise (in addition to other sources
of noises) that infests the interferometers
\begin{equation}
\sigma \sim (ct l_{QG}^2)^{1/3}.
\label{noise}
\end{equation}
In other words, quantum space-time effects provide another source of noise
in the interferometers and that noise is given by Eq. (\ref{noise}).  It
is customary to write the displacement noise in terms of the associated
displacement amplitude spectral density $S(f)$ of frequency $f$.  For a
frequency-band limited from below by the time of observation $t$, $\sigma$
is given in terms of $S(f)$ by\cite{radeka}
\begin{equation}
\sigma^2 = \int_{1/t}^{f_{max}}[S(f)]^2 df.
\label{spden}
\end{equation}
Now we can easily check that, for the displacement noise given by Eq.
(\ref{noise}) corresponding to our picture of space-time foam, the
associated $S(f)$ is 
\begin{equation}
S(f) \sim f^{-5/6} (c l_{QG}^2)^{1/3}.
\label{SD}
\end{equation}
In passing, we should mention that since we are considering a time scale much 
larger than the Planck time, 
we expect this formula for $S(f)$ to hold only for frequencies much smaller than the 
Planck frequency ($c/l_P$).  For consistency, this implies that if the $S(f)$ given by Eq. (\ref{SD}) is used in the integral in
Eq. (\ref{spden}), the integral should be relatively insensitive to $f_{max}$.  That is indeed the case as the small 
frequency region dominates the integral for $\sigma$.  Needless to say, to know the high frequency behavior of $S(f)$, one would need the correct theory of quantum gravity.

We can now use the existing noise-level data\cite{abram} obtained at the
Caltech 40-meter interferometer to put a bound on $l_{QG}$.  In 
particular, by
comparing Eq. (\ref{SD}) with the observed noise level of $3\times
10^{-19} {\rm mHz}^{-1/2}$ near 450 Hz, which is the lowest noise level reached
by the interferometer, we obtain the bound $l_{QG} \lsim 10^{-29}$ m which
is in accordance with our expectation $l_{QG} \sim l_P \sim 10^{-35}$ m.
The exciting news is that the "advanced phase" of LIGO\cite{abram2} is
expected to achieve a displacement noise level of less than $10^{-20}
{\rm mHz}^{-1/2}$ near 100 Hz, and this would probe $l_{QG}$ down to
$10^{-33}$
m which is almost the length scale that we expect it to be.  Moreover,
since $S(f)$ goes like $f^{-5/6}$ according to Eq. (\ref{SD}), we can look
forward to the post-LIGO/VIRGO generation of gravity-wave interferometers
for improvement by optimizing the performance at low frequencies.  As
lower frequency detection is possible only in space, we will probably need 
to wait for a decade or two for the LISA-type set-ups\cite{LISA}; but it
will be worth the wait!

We can also test the other two pictures of space-time foam by using the
gravity-wave interferometers.  The results\cite{result} are shown in the
accompanying Table where, for convenience, we have rewritten
Eq. (\ref{Wheeler}) and Eq. (\ref{AC}) respectively as $\delta l \gsim
L_{QG}$ and $\delta l \gsim (l \tilde{l}_{QG})^{1/2}$.  We expect both
$L_{QG}$ and $\tilde{l}_{QG}$ to be of order $l_P \sim 10^{-35}$ m.  Note 
that the amplitude spectral density for each
of the three space-time foam pictures has its own characteristic
frequency dependence.

\vspace*{1.0cm}
\begin{tabular}{||l||l|l|l||}    \hline\hline
Spacetime pictures with $\delta l \gsim$ & $L_{QG}$ & $(l l_{QG}^2)^{1/3}$ & $(l
\tilde{l}_{QG})^{1/2}$ \\ \hline
Metric fluctuations with $\delta g \gsim$ & $\frac{L_{QG}}{l}$ &
$(\frac{l_{QG}}{l})^{2/3}$ & $(\frac{\tilde{l}_{QG}}{l})^{1/2}$ \\ \hline
Displacement noise $\sigma$  & $L_{QG}$  & $(ct l_{QG}^2)^{1/3}$ & $(ct \tilde{l}_{QG})^{1/2}$ \\
\hline
Amplitude spectral density  & $f^{-1/2} L_{QG}$ & $f^{-5/6} (c l_{QG}^2)^{1/3}$ & $f^{-1} (c \tilde{l}_{QG})^{1/2}$ \\ 
$S(f)$ & & & \\ \hline
Bound from 40-m  & $L_{QG} \lsim 10^{-17}$ m & $l_{QG}
\lsim 10^{-29}$ m & $\tilde{l}_{QG} \lsim 10^{-40}$ m \\ 
interferometer & & & \\ \hline
Advanced phase of LIGO  & $L_{QG}$ to $10^{-19}$ m & $l_{QG}$ to
$10^{-33}$ m & $\tilde{l}_{QG}$ to $10^{-45}$ m \\ 
probes & & & \\ \hline
Present status & hard to check & waiting eagerly & ruled
out? \\ \hline\hline 
\end{tabular}
\medskip

\section{Conclusions}

As the last column of the accompanying Table shows, the existing noise-level data obtained 
at the Caltech 40-m interferometer have already excluded all values of
$\tilde{l}_{QG}$ down to $10^{-40}$ m, five orders of magnitude smaller 
than the Planck length.  Thus, the third picture of space-time
foam appears to be in serious trouble, if not aleady ruled out.  It is 
interesting to reflect that, until
recently, no one would have dreamed that there is a way to rule out a
space-time foam model that predicts a mere $10^{-4}$ m uncertainty on a
scale of the whole observable universe.  Now, even the Planck scale is no
longer regarded as so prohibitively small that quantum gravity cannot 
be probed by modern laser interferometry.

On the other hand, the Table also shows that the quantum space-time effects 
predicted by the canonical picture of space-time foam (corresponding to 
fluctuations given by Eq. (\ref{Wheeler}) in space-time measurements) are
still far too small to be measured by interferometry technologies currently 
available or 
imaginable.  Even the advanced phase of LIGO can probe $L_{QG}$ only
down to $10^{-19}$ m, some 16 orders away from the expected scale of
Planck length.  Waiting for the confirmation of the canonical space-time
foam picture with the techniques of interferometry is like waiting for
Godot in Beckett's play --- the waiting may never end.

Finally, here is the exciting news: modern gravity-wave interferometers are 
within striking distance of testing the space-time foam picture proposed by 
us\cite{nvd1,nvd2,karol}. Incredibly, the
advanced phase of LIGO will probe $l_{QG}$ down to $10^{-33}$ m.  We can
expect even more stringent bounds on $l_{QG}$ with future 
LISA-type projects.\cite{LISA}
According to our space-time foam picture, a noise-level corresponding to
the associated amplitude spectral density given by Eq. (\ref{SD}) with
$l_{QG}$ of the order of Planck length, should be left in the read-out of
an interferometer even after all classical-physics and ordinary 
quantum-mechanics noise sources have been eliminated.  That noise is
an intrinsic consequence of quantum gravity.  
If and when that noise is
detected, we will have successfully taken a glimpse at the very fabric of
space-time at very short distance scales.  Eagerly we wait to catch that 
faint echo from space-time quantum fluctuations.

\bigskip

\begin{center}
{\bf Acknowledgments}\\
\end{center}

We thank G. Amelino-Camelia for a useful correspondence.  This work was supported in part by the Department of Energy and by the Bahnson Fund of the University of North Carolina at Chapel Hill.

One of us (YJN) gave a seminar on the topic of space-time measurements to
a very receptive audience at the University of Connecticut in the fall of
1993.  In the audience was Prof. Kurt Haller who probably did raise
the question: How can we test the uncertainty expressed in Eq.
(\ref{ineq5})?  At that time we had no concrete and
practical idea.  Now, five and half years later, we are glad to
report to Prof. Haller that there is a way to do it.  This article is dedicated to him to celebrate his seventieth birthday.


\begin{references}
\bibitem{twoforms}
Still we wonder if quantum mechanics and general relativity are merely two forms of a more general structure.
\bibitem{LISA}
K. Danzmann, Class. Quant. Grav. {\bf 13}, A247 (1996).  Here, we are
hoping that the gain by going to lower frequencies is not offset by other
factors such as a much larger arm length of the interferometer.
\bibitem{result}
For the space-time foam model given by Eq. (\ref{Wheeler}), we have 
neglected a small (logarithmic) t-dependent correction in the $\sigma$ 
corresponding to the $S(f)$ shown in the Table.
\end{references}
\end{document}